\documentclass[twocolumn,showpacs,preprintnumbers,amsmath,amssymb,prb]{revtex4}

\usepackage{graphicx}
\usepackage{bm}

\newcommand{\vv}[1]{\mbox{\boldmath{$#1$}}}
\newcommand{\vs}[1]{\mbox{\footnotesize \boldmath{$#1$}}}
\newcommand{\avg}[1]{\left\langle {#1} \right\rangle}

\begin{document}
\title{
Nearly Perfect Single-Channel Conduction in Disordered Armchair Nanoribbons
}
\author{Masayuki Yamamoto$^1$}
\author{Yositake Takane$^1$}
\author{Katsunori Wakabayashi$^{1,2}$}
\affiliation{$^1$Department of Quantum Matter, AdSM, Hiroshima University,
Higashi-Hiroshima 739-8530, Japan
}
\affiliation{$^2$PRESTO, Japan Science and Technology Agency(JST), Kawaguchi
332-0012, Japan}
\date{\today}

\begin{abstract}
The low-energy spectrum of graphene nanoribbons with
armchair edges (armchair nanoribbons) is described as 
the superposition of two non-equivalent Dirac points of graphene. 
In spite of the lack of well-separated two 
valley structures, the single-channel transport subjected to
long-ranged impurities is nearly perfectly conducting,
where the backward scattering matrix elements in the lowest order vanish 
as a manifestation of internal phase structures of the wavefunction.
For multi-channel energy regime, however, the conventional exponential
decay of the averaged conductance occurs.
Since the inter-valley scattering is not completely absent, 
armchair nanoribbons can be classified into orthogonal universality
class irrespective of the range of impurities.
The nearly perfect single-channel conduction dominates
the low-energy electronic transport in rather narrow nanorribbons.  
\end{abstract}

\maketitle

\section{Introduction}
Graphene is the first true  two-dimensional (2D) material~\cite{novoselov}.
Due to the honeycomb lattice structure of sp$^2$ carbon,
the $\pi$ electronic states near the Fermi energy behave as
the massless Dirac fermions.
This leads to many nontrivial properties of graphene
such as half-integer quantum Hall effect~\cite{qhe}. 
The valence and conduction bands touch conically at two non-equivalent
Dirac points, called $\bm{K_+}$ and $\bm{K_-}$ points,
which possess opposite chirality~\cite{chiral}. 
In graphene, the presence of edges can have strong
implications for the electronic band structure of $\pi$
electrons~\cite{peculiar,nakada,prb}. 
Graphene nanoribbons (GNR) with zigzag edges are known
to have partial flat bands near the Fermi energy due to the
edge localized states. The electronic structures of nanoribbons
with armchair edges crucially depend on the ribbon
width~\cite{peculiar,nakada,prb,son}. 
Recent rapid progress of experiments confirmed the edge-dependent
electronic states of graphene using scanning tunneling
microscope~\cite{niimi,kobayashi}, 
and also succeeded in creating 
GNR using lithographic~\cite{han} or chemical techniques~\cite{li}.

GNR displays unusual electronic transport properties,
in apparent conflict with the common belief that 
1D systems are generally subject to  Anderson localization. Indeed it was
demonstrated that nanoribbons with zigzag edges (zigzag nanoribbons)
with long-ranged impurities 
possess one perfectly conducting channel (PCC), {\it i.e.} the absence of
Anderson localization~\cite{pcc,pcc2}. 
Since in zigzag nanoribbons the propagating modes in each valley contain
a single chiral mode 
originating from edge states, a single PCC emerges, 
associated with such a chiral mode,
if the impurity scattering does not connect the two valleys, {\it i.e.} for
long-ranged impurities (LRI). 
\begin{figure}[ht]
\begin{center}
\includegraphics[scale=0.4]{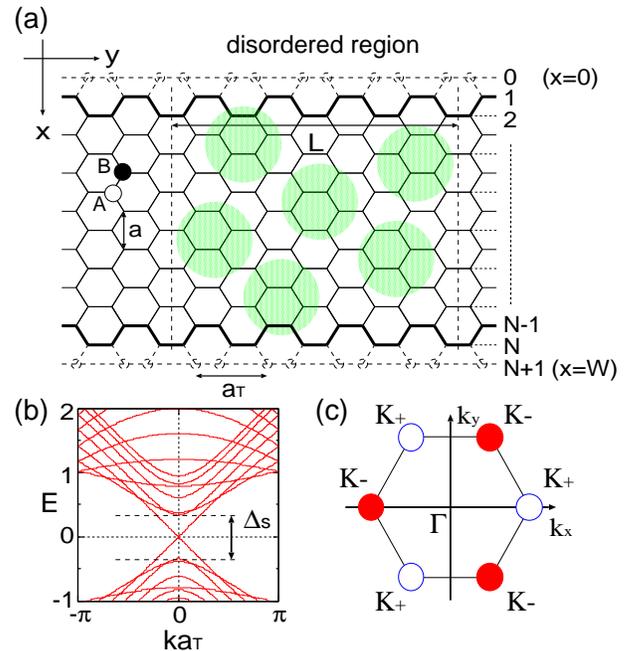}
\caption{
(a) Structure of graphene armchair nanoribbon.
The area with the length $L$
represents the disordered region with randomly distributed impurities.
(b) Energy dispersion of armchair ribbon with $N=14$.
The energy range for single-channel transport is described by $\Delta_s$. 
(c) 1st Brillouin Zone of graphene.
}
\label{fig1}
\end{center}
\end{figure}

\begin{figure*}
\includegraphics[width=0.95\linewidth]{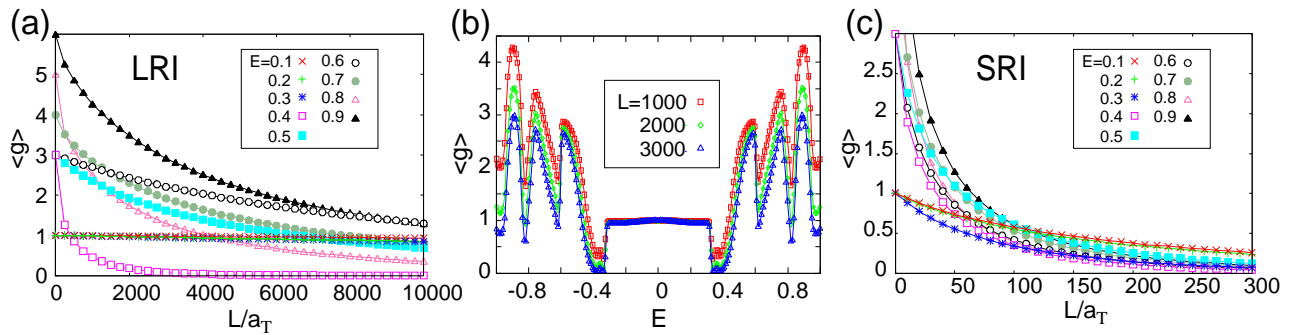}
\caption{(a) Average conductance $\avg{g}$ 
as a function of the ribbon length $L$
in the presence of long-ranged impurities (LRI)
for several different Fermi energies $E$. 
Conductance is almost unaffected by impurities for single-channel
transport ($E=0.1, 0.2 \,\,{\rm and}\,\, 0.3$) while it shows a conventional
exponential decay for multi-channel transport ($E \ge 0.4$).
Here, $N=14$, $n_{\rm imp.}=0.1$ and $d/a=1.5$. 
Ensemble average is taken over $10^4$ samples.
(b) The Fermi energy dependence of $\langle g\rangle$ for LRI. 
(c) The same as (a) for short-ranged impurities (SRI). 
Here, $N=14$, $n_{\rm imp.}=0.1$ and $d/a=0.1$. 
}
\label{fig2}
\end{figure*}

In this paper, we show that the single-channel transport in the
disordered armchair nanoribbons subjected to the 
long-ranged impurities is nearly perfectly conducting in spite of the
lack of well-separated two valley structures. 
The origin of the nearly perfect conduction is the cancellation of 
the backward scattering matrix elements in the lowest order due to 
the manifestation of internal phase structures of the wavefunction.
For multi-channel energy regime, however, the conventional exponential
decay of the averaged conductance occurs.
Since the inter-valley scattering is not completely absent, the 
disordered armchair nanoribbons can be classified into orthogonal class.
The nearly perfectly conducting effect dominates
the low-energy electronic transport properties in rather narrow nanorribbons.  

The paper is organized as follows:
In Sec.\,\ref{sec2}, the tight-binding model used in our numerical simulation is explained.
We also briefly review the electronic states of the low-energy single channel mode
in armchair nanoribbons by $\vv{k\cdot p}$ scheme. 
In Sec.\,\ref{sec3}, we present the numerical results indicating the nearly perfect single-channel conduction.
This property is then explained by $T$-matrix analysis.
Symmetry consideration is also given in this section.
Finally we summarize our work in Sec.\,\ref{sec4}.

\section{Electronic states of armchair nanoribbons \label{sec2}}
\subsection{Tight-binding model}
We describe the electronic states of graphene nanoribbons with armchair
edges by the tight-binding model
\begin{equation}
H = 
\sum_{\avg{i,j}} \gamma_{ij} c_i^{\dagger} c_j
+ \sum_i V_i c_i^{\dagger}c_i, 
\label{hamil}
\end{equation}
where 
$c_i (c_i^{\dagger})$ denotes the creation (annihilation) operator
of an $\pi$-electron on the site $i$ neglecting the spin degree of freedom.
$\gamma_{ij} = -1$ if $i$ and $j$ are nearest neighbors, and 
$0$ otherwise. 
In the following we will
also apply magnetic fields perpendicular to the 
graphite plane which are incorporated via the Peierls phase: 
 $\gamma_{ij}\rightarrow\gamma_{ij}\exp\left[
i2\pi(e/ch)\int_i^jd\bm{l\cdot A}\right]$, where $\bm{A}$
is the vector potential. 
The second term in Eq.\,(\ref{hamil}) represents
the impurity potential, $V_i=V(\bm{r}_i)$,
at position $\bm{r}_i$. 

In Fig.\,\ref{fig1}(a), the schematic figure of armchair
ribbons is depicted. The ribbon width $N$ is defined by the number of
zigzag kinks in the transverse direction. 
The armchair ribbon can be metallic if $N=3m-1$ ($m$: integer number)
as shown in Fig.\,\ref{fig1}(b), 
otherwise semiconducting. 
The disordered sample region with the length $L$ 
is attached to two reservoirs via semi-infinite ideal regions.

We assume that impurities are randomly distributed
with density $n_{\rm imp.}$.
Each impurity potential has the Gaussian form of range $d$,
\begin{equation}
V_i=
V(\vv{r}_i)= \sum_{\vs{r}_0 {\rm (random)}}
u \exp \left( - \frac{|\vv{r}_i-\vv{r}_0|^2}{d^2} \right)\,,
\end{equation}
where the strength $u$ is uniformly distributed within the range
$|u| \le u_M$. 
Here $u_M$ satisfies the normalization condition:
$u_M \sum_{\vv{r}_i}^{\rm (full space)} 
\exp(-\vv{r}_i^2/d^2)/(\sqrt{3}/2)=u_0$.
In this work,
we set $n_{\rm imp.}=0.1$, $u_0=1.0$ and 
$d/a=1.5$ for LRI and $d/a=0.1$ for short-ranged impurities (SRI).

\subsection{Low-energy single channel mode}
Here we briefly review the relation between the low-energy electronic
states of armchair nanoribbons and the Dirac spectrum of graphene. 
The electronic states near the Dirac point in graphene can be described
by the massless Dirac Hamiltonian
\begin{equation}
\hat{H}_0 = \tilde{\gamma}
\left[
\hat{k}_x(\sigma^x\otimes\tau^0)
-\hat{k}_y(\sigma^y\otimes\tau^z)
\right],
\end{equation}
acting on the 4-component pseudospinor envelope functions
$\bm{F}(\bm{r}) = 
\left[
F_{A}^{+}({\bm r}), F_{B}^{+}({\bm r}), F_{A}^{-}({\bm r}), 
F_{B}^{-}({\bm r}) \right]$,
which characterize the wave functions on the two crystalline sublattices
($A$ and $B$) for the two non-equivalent Dirac points (valleys)
$\bm{K_\pm}$ shown in Fig.\,\ref{fig1}(c).
The corresponding wave vector for the $\bm{K_+}$ point is 
$\bm{K} = (2\pi/a)(2/3,0)$, 
and that for the $\bm{K_-}$ point is $-\bm{K}$. 
We have defined the amplitude of wavefunction at ${\bm R}_A$ and
 that at ${\bm R}_{B}$ as 
$\psi_{A}({\bm R}_A)
 =  {\rm e}^{{\rm i}{\bm K}\cdot {\bm R}_{A}} F_{A}^{+}({\bm R}_{A})
    + {\rm e}^{-{\rm i}{\bm K}\cdot {\bm R}_{A}} F_{A}^{-}({\bm R}_{A})$
and
$  \psi_{B}({\bm R}_B)
 =  {\rm e}^{{\rm i}{\bm K}\cdot {\bm R}_{B}} F_{B}^{+}({\bm R}_{B})
 - {\rm e}^{-{\rm i}{\bm K}\cdot {\bm R}_{B}} F_{B}^{-}({\bm R}_{B}),$
respectively. 
Here, ${\bm R}_A$ (${\bm R}_{B}$) is the coordinate of an arbitrary
$\bm{A}$($\bm{B}$) sublattice site. 
Here $\tilde{\gamma}$ is the band parameter, $\hat{k}_x$($\hat{k}_y$) are
wave number operators, and $\tau^0$ is the $2\times 2$ identity matrix. 
Pauli matrices $\sigma^{x,y,z}$ act on the sublattice
space ($A$, $B$), while $\tau^{x,y,z}$ on the valley space
($\bm{K}_\pm$). 
The boundary condition for armchair nanoribbons~\cite{brey} can
be written as
\begin{align}
  & [F_{A}^{+}(x,y) + F_{A}^{-}(x,y)]|_{x=0,W} = 0,
     \\
  & [F_{B}^{+}(x,y) - F_{B}^{-}(x,y)]|_{x=0,W} = 0. 
\end{align}
Since this boundary condition projects
$\bm{K_+}$ and $\bm{K_-}$ states into $\Gamma$ point in
the first Brillouin Zone as seen in Fig.\,\ref{fig1}(c), 
the low-energy states for armchair nanoribbons are 
the superposition of $\bm{K_+}$ and $\bm{K_-}$ states. 
If the ribbon width $W$ satisfies the condition of $W =(3/2)(N_w+1)a$
with $N_w = 0,1,2,\dots$,  
the system becomes metallic with the linear spectrum. 
The corresponding energy is given by
\begin{align}
 \epsilon_{n,k,s} = s \tilde{\gamma} \sqrt{\kappa_{n}^{2}+k^{2}}, 
\label{eigenenergy}
\end{align}
where $\kappa_{n} = \frac{2\pi n}{3(N_w+1)a}$, 
$n = 0, \pm 1,\pm 2, \dots$ and $s = \pm$.
The $n=0$ mode is the lowest linear subband for metallic armchair
ribbons.
The energy gap ($\Delta_s$) to first parabolic subband of $n=1$ is given
as 
\begin{equation}
\Delta_s = 4\pi\tilde{\gamma}/3(N_w+1)a,
\end{equation}
which is inversely proportional to ribbon width.
It should be noted that small energy gap can be acquired due to the
Peierls distortion for half-filling at low temperatures~\cite{igami,son}, but
such effect is not relevant for single-channel transport in the doped energy
regime. 
\begin{figure}[ht]
\begin{center}
\includegraphics[width=\linewidth]{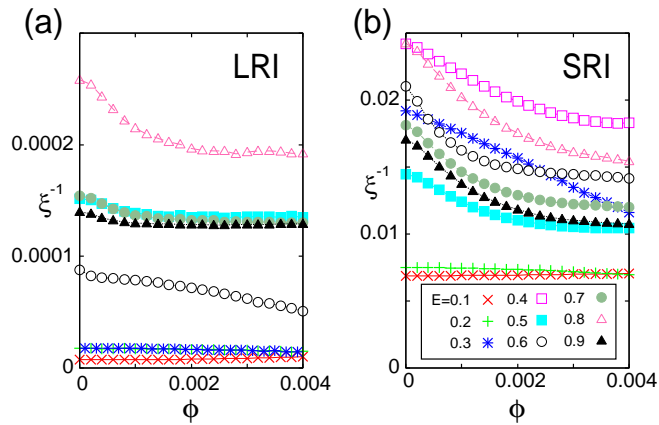}
\caption{
Inverse of localization length $\xi^{-1}$ as a function of the magnetic
flux $\phi$ through a hexagon ring measured in units of $ch/e$, 
in the presence of (a) LRI and (b) SRI for different Fermi energies $E$.
In (a), $\xi^{-1}$ for $E=0.4$ is omitted since its value is much larger than others.
\label{fig3}}
\end{center}
\end{figure}

\section{Electronic transport properties of disordered nanoribbons \label{sec3}}
\subsection{Numerical simulation}
Now we turn to the discussion of the electronic transport properties of
disordered nanoribbons. 
We evaluate the dimensionless conductance by using the Landauer formula,
$g(E) = {\rm Tr} (\vv{t}^{\dagger} \vv{t})$. 
Here the transmission matrix $\bm{t}(E)$ for
disordered system is calculated by using the recursive Green function
method~\cite{gf}. 

Figure \ref{fig2}(a) shows the averaged conductance $\avg{g}$ as a function of 
the ribbon length $L$ in the presence of LRI for several different Fermi energies $E$.
As we can clearly see, the averaged conductance subjected to LRI in the 
single-channel transport ($E = 0.1, 0.2 \,\,{\rm and}\,\, 0.3$) is
nearly equal to one even in the long wire regime. 
This result is contrary to our expectation that electrons are scattered
even by LRI, 
since wavefunctions at $\bm{K_+}$ and $\bm{K_-}$ points are mixed in armchair ribbons.
For multi-channel transport ($E \ge 0.4$), the conductance shows a conventional decay.
The robustness of single-channel transport can be clearly viewed from the 
Fermi energy dependence of conductance
for several different ribbon lengths $L$ as shown in Fig.\,\ref{fig2}(b).
It should be noted that the energy dependence in the vicinity of $E=0$
is quite different from that in zigzag nanoribbons.
The conductance decays rapidly due to the finite ribbon width
effect in zigzag ribbons~\cite{pcc} while
the conductance around $E=0$ remains unity 
in armchair ribbons (Fig.\,\ref{fig2}(b)).

Now let us see the effect of short-ranged impurities (SRI).
Figure \ref{fig2}(c) shows the
average conductance $\avg{g}$ as a function of 
the ribbon length $L$ in the presence of SRI
for several different Fermi energies $E$.
In this case, the conductance decays exponentially even for single-channel transport.
This result is similar to that previously obtained in zigzag ribbons. 
However, the rate of decay in the low-energy single-channel regime
($E=0.1 \,\,{\rm and} \,\,0.2$) is slower than that for multi-channel
transport regime ($E \ge 0.4$) in this case.  
Similar results are obtained in Ref.~\cite{lin}, but in which 
only short-ranged disorder at edge of ribbon is considered. 

\subsection{$T$-matrix analysis}
The absence of localization in the single-channel region can be 
understood from the Dirac equation including the
impurity potential term $\hat{U}_{\rm imp}$ with armchair edge boundary. 
To consider the amplitude of backward scattering, we introduce
the $T$-matrix defined as
\begin{align}
  T = \hat{U}_{\rm imp}
      + \hat{U}_{\rm imp}\frac{1}{E - \hat{H}_{0}}\hat{U}_{\rm imp}
      + \cdots .
\end{align}
According to Ref.\,\cite{ando.nakanishi},
$\hat{U}_{\rm imp}$ is written as
\begin{align}
    \hat{U}_{\rm imp}
  = \left( \begin{array}{cccc}
               u_{A}({\bm r}) & 0 & {u'}_{A}({\bm r}) & 0 \\
               0 & u_{B}({\bm r}) & 0 & -{u'}_{B}({\bm r}) \\
               {u'}_{A}({\bm r})^{*} & 0 & u_{A}({\bm r}) & 0 \\
               0 & -{u'}_{B}({\bm r})^{*} & 0 & u_{B}({\bm r}) \\
           \end{array}
    \right),
\end{align}
with 
\begin{align}
 u_{X}({\bm r}) & = \sum_{{\bm R}_{X}} g \left(\bm{r}-{\bm R}_{X}\right)
                     \tilde{u}_X\left({\bm R}_{X}\right)\,,
     \\
    \label{eq:u_inter}
 {u'}_{X}({\bm r}) & = \sum_{{\bm R}_{X}} g \left(\bm{r}-{\bm R}_{X}\right)
                     {\rm e}^{-{\rm i}2 {\bm K}\cdot {\bm R}_{X}}
                     \tilde{u}_X\left({\bm R}_{X}\right)\,,
\end{align}
where $\tilde{u}_{X}(\bm{R}_{X})$ is the local potential due to impurities
for $X = A$ or $B$.
Here $g(\bm{R})$
with the normalization condition of $\sum_{\bm{R}}g(\bm{R})=1$ 
is the real function which has an appreciable amplitude in
the region where $\rvert\bm{R}\rvert$ is smaller than a few times of the 
lattice constant, and decays rapidly with increasing $\rvert\bm{R}\rvert$. 
If only the LRI are present, we can approximate
$u_{A}({\bm r}) = u_{B}({\bm r}) \equiv u({\bm r})$ and
${u'}_{A}({\bm r}) = {u'}_{B}({\bm r}) \equiv {u'}({\bm r})$.
In the case of carbon nanotubes and zigzag nanoribbons, ${u'}_{X}({\bm r})$
vanishes after the summation over $\bm{R}_{X}$ in Eq.~(\ref{eq:u_inter})
since the phase factor ${\rm e}^{-{\rm i}2 {\bm K}\cdot {\bm R}_{X}}$
strongly oscillates in the $x$-direction.
%
However, this cancellation is not complete in an armchair nanoribbon because
the averaging over the $x$-direction is restricted to the finite width of $W$.
This means that we cannot neglect the contribution 
from scatterers particularly in the vicinity of the edges
to ${u'}_{X}({\bm r})$.
Although ${u'}_{X}({\bm r})$ becomes small after the summation,
the symmetry of system changes for ${u'}_{X}({\bm r}) \ne 0$
as we will see in the next subsection.

Now we evaluate the matrix elements of $\hat{U}_{\rm imp}$ for
the eigenstate $\rvert n,k,s \rangle$ with the eigen energy of
Eq.\,(\ref{eigenenergy}) which can be written as 
\begin{align}
 \rvert n,k,s \rangle
  = \frac{1}{\sqrt{4WL}}
    \left(
    \begin{array}{l}
      \left( \begin{array}{c}
                 s \\
                 {\rm e}^{-{\rm i}\theta(n,k)}
             \end{array}
      \right) {\rm e}^{{\rm i}\kappa_{n}x}
      \\
      \left( \begin{array}{c}
                 -s \\
                {\rm e}^{-{\rm i}\theta(n,k)}
             \end{array}
      \right) {\rm e}^{-{\rm i}\kappa_{n}x}
    \end{array}
    \right)  {\rm e}^{{\rm i}ky}\,,
\label{wf}
\end{align}
with the phase factor
\begin{align}
{\rm e}^{-{\rm i}\theta(n,k)} 
   = \frac{\kappa_{n}-{\rm i}k}{\sqrt{\kappa_{n}^{2}+k^{2}}}\,. 
\end{align}
Here it should be noted that the phase structure in Eq.\,(\ref{wf}) is 
different between $\bm{K_+}$ and $\bm{K_-}$ states, 
and this internal phase structures are critical for the
scattering matrix elements of armchair nanoribbons
as we discuss in the following.
Using the above expression, we can obtain the 
scattering matrix element
\begin{eqnarray}
     \label{eq:U_matel}
\langle n,k,s \lvert \hat{U}_{\rm imp} \rvert n',k',s' \rangle 
&=& \left( ss' + {\rm e}^{{\rm i}(\theta(n,k)-\theta(n',k'))} \right) \nonumber \\
&\times&  V\left(n,k; n^\prime,k^\prime \right)\,,
\label{eq14}
\end{eqnarray}
with
\begin{eqnarray}
 V\left(n,k; n', k' \right)
&=& \frac{1}{4WL}\int_{0}^{W}{\rm d}x \int_{0}^{L}{\rm d}y \,
  {\rm e}^{-{\rm i}(k-k')y}\nonumber\\
&\times&
 \left[  u(\bm{r}) \left( {\rm e}^{-{\rm i}(\kappa_{n}-\kappa_{n'})x}
                             + {\rm c.c.} \right) \right.\nonumber\\
 &-& \left.\left( u'(\bm{r}){\rm e}^{-{\rm i}(\kappa_{n}+\kappa_{n'})x}
                    + {\rm c.c.} \right) 
    \right] .
\label{eq15}
\end{eqnarray}
It should be emphasized that Eq.\,(\ref{eq14}) has the same form as that
obtained for carbon nanotubes without inter-valley scattering ($u'_X(\bm{r}) = 0$)\,\cite{ando.nakanishi}.
Interestingly, in spite of the fact that armchair nanoribbons inevitably suffer from the inter-valley
scattering due to the armchair edges ($u'_X(\bm{r}) \ne 0$),
we can express the matrix element for the backward scattering as Eq.\,(\ref{eq14})
by including $u'_X(\bm{r})$ into $V(n, k; n', k')$ in Eq.\,(\ref{eq15}).
This is due to the different phase structure between $\bm{K_+}$ and $\bm{K_-}$
in Eq.\,(\ref{wf}).

We focus on the single-channel regime where only the lowest subband
with $n = 0$ crosses the Fermi level.
From Eq.\,(\ref{eq:U_matel}), 
the scattering amplitude from the propagating state 
$\rvert 0,k,s\rangle$ to its backward state
$\rvert 0,-k,s\rangle$ in the single-channel mode becomes
identically zero, {\it i.e.} 
\begin{align}
  \langle 0,-k,s \lvert \hat{U}_{\rm imp} \rvert 0,k,s \rangle = 0 .
\end{align}
Thus, since the lowest backward scattering matrix element of $T$-matrix
vanishes, the decay of $\langle g \rangle$ in the single channel energy
regime is extremely slow as a function of the ribbon length as we have
seen in Fig.\,\ref{fig2}.
However, the back-scattering amplitude in the second and much higher order
does not vanish.
Hence the single-channel conduction is not exactly perfect like carbon nanotubes\,\cite{ando.nakanishi},
but {\it nearly} perfect in armchair nanoribbons.

The present results of nearly perfect single-channel transport might be similar to 
those obtained in carbon nanotubes
by solving the Boltzmann transport equation, which is valid for incoherent systems
in the absence of inter-valley scattering\,\cite{ando.suzuura}.
However, our results are for the coherent system with inter-valley scattering by armchair edge
and their physical mechanism is different.

\subsection{Symmetry consideration}
Now we give a symmetry consideration to disordered graphene and graphene
nanoribbons. If the inter-valley scattering is absent, 
{\it i.e.} $u'_X(\bm{r})= 0$, 
the Hamiltonian $\hat{H}_{0} + \hat{U}_{\rm imp}$ 
becomes invariant under the transformation of
$\mathcal{S} = -{\rm i}\left(\sigma^{y}\otimes \tau^{0}\right)C$,
where $C$ is the complex-conjugate operator.
This operation corresponds to the special time-reversal operation for pseudospins within
each valley, and supports that the system has the symplectic symmetry.
However, in the presence of inter-valley scattering due to SRI, 
the invariance under $\mathcal{S}$ is broken.
In this case, the time reversal symmetry across two valleys described by
the operator $\mathcal{T}= \left(\sigma^z\otimes\tau^x\right)C$
becomes relevant, which indicates orthogonal universality class.
Thus as noted in Ref.~\cite{suzuura.prl}, 
graphene with LRI belongs to symplectic symmetry, but that with SRI
belongs to orthogonal symmetry. 

However, in the disordered armchair ribbons,
the special time-reversal symmetry within each valley is broken
even in the case of LRI.
This is because $u'_X(\bm{r}) \ne 0$ as we have seen in the previous subsection.
Thus, irrespective of the range of impurities,
the armchair ribbons are classified into orthogonal universality class. 
Actually, the application of magnetic field shows
rather strong magnetic field dependence on the inverse localization
length in the regime of the weak magnetic field 
for both LRI and SRI cases (Fig.\,\ref{fig3}).
This is consistent with the behavior of orthogonal universality class. 
Here, the inverse localization length is evaluated by identifying
$\exp\langle\ln g\rangle=\exp(-L/\xi)$. 
Since the disordered zigzag nanoribbons are classified
into unitary class for LRI but orthogonal class for SRI~\cite{pcc}, 
it should be noted that the universality crossover in nanographene
system can occur not only 
due to the range of impurities but also due to the edge boundary
conditions.

\section{Summary \label{sec4}}
In this work, we have numerically investigated the electronic transport
in disordered armchair nanoribbons in the presence of short- and
long-ranged impurities. 
In spite of the lack of well-separated two 
valley structures, the single-channel transport subjected to
long-ranged impurities shows nearly perfect transmission, 
where the backward scattering matrix elements in the lowest order vanish 
as a manifestation of internal phase structures of the wavefunction.
These results are contrast with the mechanism of perfectly conducting
channel in disordered zigzag nanoribbons and metallic nanotubes
where the well separation between two non-equivalent Dirac points is
essential to suppress the inter-valley scattering. 
Within the Born approximation, this 
cancellation is satisfied even for SRI, which can be clearly seen in 
Fig.\,\ref{fig2}(b). The dependency of conductance on the Fermi
energy confirms our calculation about the
difference between single- and multi-channel transport properties. 
The symmetry consideration classifies the armchair nanoribbons 
into orthogonal class. 

\section*{Acknowledgement}
This work was financially supported by a Grand-in-Aid for Scientific
Research from the MEXT and the JSPS (Nos. 19710082, 19310094 and 20001006). 


\end{document}